# The Social Sciences Interdisciplinarity for Astronomy and Astrophysics - Lessons from the History of NASA and Related Fields

Principal Author:
Anamaria Berea (Blue Marble Space Institute of Science and Complex Adaptive Systems Lab, University of Central Florida), anamaria@bmsis.org

Co-Authors:
Kathryn Denning (York University)
Monica Vidaurri (NASA GSFC)
Kimberly Arcand (Chandra X-ray Observatory Center, Center for Astrophysics | Harvard & Smithsonian)
Michael P. Oman-Reagan (Memorial University)
Jillian Bellovary (CUNY - Queensborough Community College)
Arsev Umur Aydinoglu (Middle East Technical University)
Mark Lupisella (NASA Goddard Space Flight Center)

Co-Signers:
Chuanfei Dong (Princeton University)
Douglas Vakoch (METI)
Lisa Messeri (Yale University)
Steven Dick (Former NASA Chief Historian)
Chanda Prescod-Weinstein (University of New Hampshire)
Beth Laura O'Leary (New Mexico State University)
Beck E. Strauss (*Independent researcher*)
Shawn D. Domagal-Goldman (NASA Goddard Space Flight Center)
Linda Billings (National Institute of Aerospace)
Jason T. Wright (Center for Exoplanets and Habitable Worlds)
Lauren M. Chambers (Space Telescope Science Institute)
Gourav Khullar (University of Chicago)
Leah Fulmer (University of Washington)

**I. Executive Summary:**

**In this paper we showcase the importance of understanding and measuring interdisciplinarity and other *-disciplinarity* concepts for all scientists, the role social sciences have historically played in NASA research and missions, the sparsity of social science interdisciplinarity in space and planetary sciences, including astronomy and astrophysics, while there is an imperative necessity for it, and the example of interdisciplinarity between social sciences and astrobiology. Ultimately we give voice to the scientists across all fields with respect to their needs, aspirations and experiences in their interdisciplinary work with social sciences through an ad-hoc survey we conducted within the Astro2020 Decadal Survey scientific community.**

**II. Key Points:**

**1. The many forms and functions of social sciences**
Social science comes in many different forms and accordingly there are multiple lines of connection with NASA and space sciences. Social sciences disciplines include psychology, anthropology, space archaeology, sociology, economics, communication, and many others. Some of the studies that can immediately pertain to interdisciplinarity with astronomy, astrophysics and beyond can be classified in:
1. Meta-studies of the sciences, scientists and scientific process, such as: the studies of scientists and scientific process themselves, the study of knowledge transmission across scientific communities, the study of education systems and effectiveness of STEM (science, technology, engineering and math) education, etc.
2. Methodological and epistemological studies, where methods from either disciplines (social sciences - astronomy/astrophysics and related) are being used for scientific research.
3. The past and current research on social phenomena that has implications for astronomy, astrophysics and related fields, such as research on global systems, human behavior at micro and macro levels, small and global communities, civilizations, financial systems, instability, system failures, complex human-technological-natural systems, and others.

Social scientists have made and can make numerous contributions to NASA endeavours. Physicists, bioscientists, astronomers and astrophysicists are often asked to speak about human issues, for public outreach, yet they often do not have the expertise to touch on such issues. Such researchers can do it better, however, with input from social scientists, particularly working as a team. Many successful research leaders also have cross-training such as Masters in public administration or MBA studies, which of course includes work from the social sciences. The Master of Space Studies program at the International Space University (ISU) explicitly incorporates interdisciplinary training, including culture. Similarly, some biological and physical scientists can also be social scientists. It is common enough to do a first degree in science and then segue to social science, or to combine social science methods with other science expertise. Many challenges in science today also involve large teams with coordination challenges. This was recognized early on by the NASA Astrobiology Institute (NAI) (Blumberg, 2004) and was the reason why Professor Blumberg hired an anthropologist near the time of program conception to help design the collaboration process for early NAI endeavors (Pilcher, 2015).

Other examples of social scientists working with engineers and physicists include psychologists that have worked and continue to work on the human spaceflight program or the human factors specialists that are working on problems of the functioning of scientific teams and on the science of teams, as well as behavioural approaches drawing from statistical and case studies (e.g. Harrison 2002). Additionally, questions about the evolution of intelligence (human and otherwise) are becoming integrated into astrobiology, with the larger framework of the evolution of human societies often invoked by other scientists. Also in the astrobiology field, a large body of work done by social scientists refers to the social implications of different types of astrobio-related work; in this respect, social scientists and humanists are engaged in questions about societal significances of detections (Denning, 2013; Denning, 2018; Funaro 2004; Harrison 1997).

A further example of social science contributions to space sciences comes from the field of anthropology and archeology, with in-depth discussions about space discourses and the framing of the scientific discourse in accurate, detailed and updated terms across disciplines (Denning & Race, 2010; Gorman, 2016; Oman-Reagan et al., 2018; Race et al., 2012). Anthropology has also significantly contributed to the imagining and designing of space colonies and also to questions of whether humans should colonize other planets (Billings, 2006, 2017, 2018; Finney 1987, 1992; Oman-Reagan, 2019).

**2. Interdisciplinarity, cross-disciplinarity, multidisciplinarity and transdisciplinarity**
Marilyn Stember's paper (1997) "Advancing the social sciences through the interdisciplinary enterprise" offers the following definitions of different levels of disciplinarity:
- Intradisciplinary: working within a single discipline.
- Cross-disciplinary: viewing one discipline from the perspective of another.

- Multidisciplinary: people from different disciplines working together, each drawing on their disciplinary knowledge.
- Interdisciplinary: integrating knowledge and methods from different disciplines, using a real synthesis of approaches.
- Transdisciplinary: creating a unity of intellectual frameworks beyond the disciplinary perspectives. (Stember, 1990)

For the purpose of this paper, we will refer to all these different nuances of science as "interdisciplinarity". In the space of planetary and space science, interdisciplinarity has often had a subjective understanding and response by various scientists. As Dr. Michael H. New of NASA pointed out, one research problem that needs to be tackled by all scientists is a common understanding across disciplines on what "interdisciplinarity" is and how we can measure it, as projects that are interdisciplinary between, for example, geochemistry and geology have a different level of crossing between disciplines than projects that include astrophysics and social science. For future interdisciplinary projects it will be useful and educational for the community to point to the correct term with respect to the theoretical and methodological mingling of disciplines. A similar comment was made by the National Research Council (NRC, 2008) in their assessment of the NASA Astrobiology Institute in which the challenges of measuring interdisciplinary research and how to advance it was discussed.

In academia there is a clear relationship among science, technology and society (STS), with academic departments and degrees that are focusing on this interdisciplinary aspect, such as "Science and Society", "Science in Society" or "Computational Social Sciences" to give only a few examples. Other academic examples include Master's and MBA degrees designed with a focus on scientists that aim to pursue careers in science policy, science management or science communication. While the practical aspects of the day to day doing of science ubiquitously show the interplay between the social sciences and other sciences, this is less obvious or prevalent in the research communities and even less so for the space sciences and planetary sciences communities.

### 3. Social Sciences and Other Sciences
#### 3.1. Social Sciences and Planetary Sciences

Some interesting interdisciplinary studies come from planetary science and importance of equity and inclusion of women in planetary missions. These were published at the Planetary Science Decadal Survey 2013-2022. The authors recommend that:

> *"Alongside scientific and technical considerations, the Planetary Science Decadal Survey should require that missions incorporate deeper consideration of the social science of spacecraft operations to maximize their missions' scientific, technical and fiscal success." (Vertesi et. al, 2009)*

Other studies on social sciences and planetary sciences refer to the fields of planetary defense and planetary protection (Billings, 2015). Studies in space economics and anthropological work related to planetary defense and planetary protection have been also recently conducted by Alissa Haddaji. These emphasize the need for sound economics and business studies in the field (Haddaji, 2019). Apart from these studies, the intersection between social sciences and planetary sciences is sparse.

#### 3.2. Social Sciences and Astrophysics and Astronomy

The authors found scant research on interdisciplinarity of social sciences with astrophysics and astronomy. One notable project was PENCIL, led by Prof. Mikaela Sundberg who carried out a series of studies from 2010 to 2012 on simulation cultures and astrophysics and the role of organizations on simulation collectives (Sundberg, 2010; Sundberg 2011; Sundberg, 2012). Ethnographic work on astronomy is an additional form of collaboration that offers opportunities to develop further interdisciplinary projects (e.g. Hoeppe 2012).

Another project of interest, although not primarily in social sciences, but towards humanities that cross into social sciences is "Aesthetics & Astronomy," a program that studies the perception of multi-wavelength astronomical imagery and the effects of the scientific and artistic choices in processing astronomical data. The team consists of a unique combination of professional astronomy communicators, astrophysicists, and aesthetics experts from the discipline of psychology. The Aesthetics & Astronomy group conducts online studies, exhibit studies, and in-person focus groups (see Smith, et al., 2011, 2015a, 2015b, 2017a, 2017b). Their research questions are designed to test such issues as:
1. How much do variations in presentation of color, explanation, and scale affect comprehension of astronomical images?
2. What are the differences between various populations (experts, novices, students) in terms of what they learn from the images?
3. What misconceptions do the non-experts have about astronomy and the images they are exposed to?

More specifically, Kimberly Kowal Arcand's research project with the University of Otago & NASA's Chandra X-ray Observatory "Putting The Stars Within Reach" investigates issues at the intersection of social science, computer science and astronomy.  One facet of this pertains to women's contributions to STEM fields, which though substantial, are often undersold and not often recognized. Parity has not been reached in women's representation numbers in STEM careers, their pay, or their rank as compared to men in the same careers (Holman, et al., 2018; Landivar, 2013; Moss-Racusin, et al., 2012; National Science Foundation, 2018; Shen, 2013).

Arcand's program focuses on improving spatial reasoning skills and STEM interest in underrepresented groups of young female learners with such a program to help build confidence and expand career interests, recognizing that those who identify as female can belong to multiple sub groups of underrepresented learners, and that issues of intersectionality cannot be separated from individuals (Crenshaw, 1991). Additionally, it is recognized that among potential cultural biases, stereotype threats, and/or issues with group and personal identities, there is not any single or simple "fix" for the issues at hand.  With that said, the research question for this project is: *What are the quantifiable effects of data-based 3D models and prints on improving spatial reasoning skills and interest (and any subsequent confidence) in STEM fields in under-represented groups, particularly of young female learners? (Arcand, et al., 2018, 2019).*

### 3.3. Social Sciences and Astrobiology

Perhaps historically the field with most prominent crossovers in social sciences has been astrobiology. One of the most influential and prestigious fellowships in astrobiology, the Blumberg chair at the Library of Congress, is specifically focused on recognizing and spurring research at the crossover of the humanities and social sciences with astrobiology. The work of the past Blumberg chairs is impressive in the depth and breadth of interdisciplinarity (https://www.loc.gov/programs/john-w-kluge-center/chairs-fellowships/chairs/blumberg-nasa-chair-in-astrobiology/). The state of the profession for technosignatures research (Wright et al., 2019) submitted for this decadal survey is a comprehensive and in-depth overview of the state of the profession describing large community efforts that cross multiple fields of study. Another overall look at astrobiology is taking a view at the culture evolution of the field (Billings, 2012). A comprehensive history of the astrobiology field and the connections of astrobiology with society have also been compiled by Doug Vakoch in a comprehensive book (Vakoch, 2013) and in two volumes from Steven J. Dick (Dick, 2015, 2018).

Astrobiology and social sciences have a long history together and multiple topics and avenues of common pursuit. One of them, as mentioned before and also in the next section, is on social impact and strategies in case of detection (Shostak and Oliver, 2000), while others are exploring both the history of the field of astrobiology and the evolution of its connections with philosophy, social sciences and humanities (Billings, 2016). Some other distinctive contributions of social sciences into astrobiology include the rise and fall of social systems and civilizations, the debates around the Anthropocene, and the theoretical models of the relationships of humans with the technology, the environment and other life forms in the Universe (Denning & Berea, 2019). The perspective of social science into astrobiology - and any other sciences, as a matter of fact - relate not only to getting new perspective on "what you do", but also on "how you do what you do" (Denning & Berea, 2019).

One of the least represented social sciences in any of these efforts is probably economics. While economics is ubiquitously needed in any research, from project proposal to the management of teams, there is virtually no interdisciplinarity between economics and the space or planetary sciences. Although economics can provide numerous insights with respect to new methodologies for understanding complex phenomena, systems failure or uncertainty, to name only a few, or theoretical advances with respect to understanding human behavior and society patterns both micro and macro, these aspects of the field are virtually unknown in the astronomy and astrophysics research (Berea, 2019).

**4. Social Sciences at NASA**
The original studies on the Space shuttle and social space science were some of the most detailed and inclusive on the role and importance of social scientists for the agency (Woodfill, 1983) and for NASA missions overall.
> *"The social science study of space technology in general provides a perspective useful to a study of the Shuttle system. During the 1960s and 1970s a variety of books, articles, studies, and research projects addressed the relation of space technology to one or several aspects of society. By 1978 the volume of activity was sufficient to prompt NASA to commission an inventory and analysis of research on space technology produced by the social science and humanities disciplines." (Woodfill, 1983)*

The study provided in-depth analyses such as on utilizing space to encourage economic and social growth or conduct impact analyses (to name only a few), and focuses on several social sciences, from economics, law, history, psychology and sociology to the future studies and the use of methods from social sciences, such as surveys, polls, simulations, scenarios and many more. The lessons from social sciences and the shuttle program have also been outlined in "Wowing the public: The shuttle as a cultural icon" (Billings, 2013) as well as in societal impacts of spaceflight (Dick, 2007).

NASA is an agency and workplace that is unique in many regards. From deep space missions to crewed launches and the potential travel to Mars, there are numerous ways that NASA can utilize social sciences, which begin very early in the science traceability matrix. Unfortunately, the current framework of integration between social and natural sciences is nearly non-existent at NASA.

Although NASA has a wide range of social science representation in media, communications, history, policy, and other external affairs, integration of social sciences in science missions across all directorates only appears in the fields of sociology and crisis management. Social scientists in these fields are available to PIs, lab chiefs, and other management in cases of hardship within a team of researchers or amongst employees, representing the full extent of an overwhelming majority of social scientist participation in science missions.

Though social science is mainly represented in the form of human capital, only common demographics such as race, gender, and disability are taken annually, highlighting a missing component that is necessary for an agency of this nature: workforce development. Social scientists are critical in determining and shaping social workplace environment, diversity and inclusion, needs of early career researchers, needs of people with disabilities, and maintaining safety and wellness of researchers. In addition, social scientists involved in science missions would modernize customs and norms of science and exploration, thus maintaining peaceful uses of space and properly communicating with and educating Congress on matters pertaining to space research. As space legislation and international customs regarding procedures in space braces for changes with respect to a growing and increasingly privatized space industry, social scientists will be the key to communicating the needs of the industry, advocating for a proactive versus a (currently) reactive space law, and bridging the gap between government and private.

The dearth of social scientists accounting for the communications, ethics, and policy, also means a lack of social scientists accounting for diversity, workplace environment, and a potential component that would prove critical to the NASA workforce: monitoring workload, growth, and the needs of early career and student researchers. Social scientists, in this case, can obtain data that will inform NASA of ways to improve student inclusion in the workplace, monitor workload, help students obtain an informed and secure career path, and ensure adequate preparedness for their careers.

In the average review panel and proposing team, there are typically no sitting members with the proper qualifications to ask or answer legal, Congressional, ethical, communicative, and international relations questions regarding the mission. This effectively does not prepare a science mission team for necessary and inevitable checkpoints and processes. One of the more famous examples of this is the James Webb Space Telescope (JWST). It was not until JWST faced a 29-month delay that NASA set up an independent review board to monitor issues regarding the delay. According to the board, major factors of the delay included human error, systems complexity, lack of experience in key areas, and excessive optimism.[1]

From this, NASA can learn two lessons: The lesson of establishing boards, liaisons, and committees to monitor social science, mission progress, and team development to catch these issues early-on, and the lesson of having social science tied to missions throughout mission traceability to facilitate progression and mitigate risk throughout mission timeline.

A study that looked at the implications of faulty decision processes that led to the Challenger disaster pinpointed the risks of not using what we currently understand about human behavior into major missions. (Vaughan, 1996). Another study examined the sociological factors for future missions based on the experience from the Apollo missions and the interviews gathered at the time (Lundquist et. al, 2011), while a sociologist working for NASA argued for the importance of methodological innovations the field can bring (Guice, 1999).

Other studies conducted by NASA for long term missions are related to the science of teams, psychology and cognitive sciences (i.e., the SCALE project). Methodologically, at the intersection between computer sciences (artificial intelligence) and social sciences is the work of William Clancey, who led the Human-Centered Computing at NASA Ames Center and whose work is cross-disciplinary among several fields of study (e.g. Clancey 2002, 2012).

An interesting study on the language of work for Mars expeditions, also from NASA Ames, emphasized the design of an effective language between scientists and engineers from the Mars Rover mission (Wales, 2005).

And finally, for the purposes of this section, NASA conducted in 2018 a Technosignatures workshop in Houston, TX, where scientists from various fields, including the social sciences, gathered for a few days to determine, among other important ideas, the importance of interdisciplinarity and inclusion of social sciences in the quest for intelligent life in the Universe (NASA Technosignatures Report, 2018). The workshop was held at the request of Rep. Lamar Smith (R-TX), but the funding was not appropriated.

**III. Recommendations:**

**5. Case Studies, Surveys and the Voices of the Scientists**
For this call for papers, we ran an ad-hoc survey with the scientists on both sides of the sciences. We received an overwhelming response of interest in the topic. Here are some examples of questions the scientists in the space and planetary sciences hope the social sciences can provide answers for. We compiled these questions from multiple iterations within the community of the scientists that are responding to the Astro2020 call.
1. What kind of things do we need to improve workforce development, diversity, collaboration, etc. that social scientists can help us with?
2. Why don't we use social scientists as often as we should?
3. What data can social scientists measure to help us move forward with the many different projects we do?
4. What kind of outreach and funding opportunities can we pursue together? Can we organize workshops or some way to connect scientists in the field with social scientists that could study it?

The community of people that have contributed to this effort and highlighted their experiences in interdisciplinary research projects comprise anthropologists, ethnographers, cosmologists, astrophysicists, astronomers, computational social scientists, economists, astrobiologists, complex systems theorists, data scientists, and biologists. Most of them have worked in interdisciplinary projects, between social sciences and other sciences. The following case studies, survey answers and opinions come from the first account of this community.

One example of an interdisciplinary multi-year project of ethnography within the planetary science community (Messeri, 2016) was the study of a number of teams at both universities and NASA to observe the day to day work to best understand both the science and sociality that goes into producing knowledge about the cosmos. The anthropologist takes seriously the "participant" part of participant observation (the idea that in order to know another community you have to be as involved as possible; not just a fly on the wall) and was fortunate to have worked with scientists who thought creatively about how the work as an anthropologist could enhance research goals. The social science participation was most seamlessly integrated by finding communication related tasks, be they public engagement or finding the right language to communicate scientific findings within an expert community. These were mutually beneficial because they allowed the close collaboration needed to do the social science research while offering skills that are often beyond the interest or aptitude of traditional scientific workers.

What can make interdisciplinary collaboration challenging is the different timescales that scientific and social scientific research operate on. The social scientist is closely in contact with the scientific community when s/he is collecting data (during participant observation). The anthropological research findings aren't articulated until well after this day to day work – after several years of processing and writing through the ethnographic material. At that point, the close connection cultivated during fieldwork has attenuated and any "findings" might not feel relevant to the scientific community as the projects originally observed and analyzed are in the past. To the social scientist, this raises an exciting challenge regarding how interdisciplinary projects can be conceived to accommodate the different time scales of the research process. Further, the challenge of the social scientist is to extract the findings that are relevant beyond the past work and be applicable moving forward.

To that end, one more challenge of interdisciplinary work is how the very different goals of research can be appreciated across the disciplines. For example, those in the planetary science community might read and perhaps even enjoy these results when published in a book, but because they are written to further anthropological knowledge (as opposed to astronomical), its applicability is not apparent. Genuine work would be needed (another interdisciplinary undertaking) to really discuss and study these research findings and think creatively about if and how they could be useful within the space sciences or NASA efforts. How can this work – that would necessarily happen after the basic research is conducted – be institutionally or financially supported? In other words, to reap the benefits of interdisciplinary work, what innovations are needed beyond conducting the research to ensure that there is an enriching return on any investment?

Another example is the perspective the social scientists have on Human Exploration, that is, how social relevance (e.g., inspiration, geopolitical, long-term survival) and meaning-making (e.g., informing worldviews, "purpose," etc.) are relevant to the social side of human exploration and are also associated with astronomy and astrophysics, such as on questions of contamination of Mars, including ethical and philosophical considerations. Cross-cutting are the perspectives on cosmology and modern science informing meaning and worldviews in a cosmic context (Dick & Lupisella, 2012; Herman, 2011; Lupisella, 2019). The interesting angle is how advanced civilizations might be engaged in large-scale, astrophysically/observationally relevant "engineering" activities, in which case, the physical large-scale impact of cultural/social civilizations would be a factor and arguably a social science kind of question (Smith et al., 2019).

Another example from the community on social science projects at or "around" NASA concerns anthropological research on "human factors," which has rarely addr about anthropological and social scientific questions, so much as it was about questions that fit better into fields Industrial Organizational Psychology, or into biological studies of bodies interacting with machines, of people and schedules, modes of work, etc. However, this focus on using social scientific disciplines as tools to enact "hard" science goals excludes a majority of the contemporary social scientific research from NASA projects. There are social scientists studying human interactions with both built and natural environments in space, social scientists studying religion in space, and others studying just about every aspect one can imagine. All of this work is producing insights that should be integrated into NASA's programs, but currently isn't.

One thing to notice in this type of interdisciplinary work is that members of the space science community often cite outdated ideas from the social sciences that have become part of popular "common sense," despite having been abandoned by the disciplines that originally proposed them (particularly certain ideas about development, advancement, evolution, etc.). A review of social scientific work included in the Astrophysics Data System demonstrated how often interdisciplinary social scientific work is excluded as well, unless it is published in a limited selection of journals, or by social scientists in included journals such as *Acta Astronautica*. However there is work beyond those traditional sources.

A resource that would particularly open up opportunities to embed more social science in their work would be the creation of a liaison position to organize social scientific research within the organization by outside researchers. The US National Park Service (NPS) has an office like this, and if a social scientist wants to do a study at a National Park, they submit a standardized application to the Park Service. As the NPS website explains: "The National Park Service (NPS) uses the web served Research Permit and Reporting System (RPRS) to administer scientific studies and collecting activities within units of the National Park System." (NPS 2019). Though still scarce overall, there are an increasing number of social scientists studying human engagements with space, and doing collaborative work with NASA scientists across disciplines, and such an office might provide a much needed resource.

### 5.1. An ad-hoc survey of the community
The following shows the questions and the summary of answers the Astro2020 community of scientists and other scientists with long experience in interdisciplinary research have kindly and candidly provided.

*1. What is interdisciplinarity for you? What do you understand to be an interdisciplinary team or interdisciplinary project? Have you been a part of interdisciplinary teams or projects? If yes, what was the main question you were trying to answer and what was your experience with this (success and failures)?*
The answers of the community on this question range from the practical experience in the field (mostly with NAI projects, the Astrobiology Primer), to understanding science as a planetary phenomenon or a more formal definition of interdisciplinarity:
> *"Multidisciplinarity draws on knowledge from different disciplines but stays within their boundaries. Interdisciplinarity analyzes, synthesizes and harmonizes links between disciplines into a coordinated and coherent whole. Transdisciplinarity integrates the natural, social and health sciences in a humanities context, and transcends their traditional boundaries." (Choi & Pak, 2006).*

The Astro2020 community also mentioned the process of interdisciplinarity as the attempt to form a common picture among all of us (24 authors, 8 editors), solidify our understanding of our own field, make us familiar with other fields, and provide us with tools to talk to a wider audience, while noticing challenges in interdisciplinarity with the humanities. Other answers included the assessment of true interdisciplinarity as incorporating perspectives, literature and methods from within and across disciplines at the earliest formulation of the research questions, working with interdisciplinary teams on questions about SETI, astrobiology, and visions of the future, while the challenges involved miscommunication, differences in literature reviews and terminology across disciplines, but also preconceptions, stereotypes, and "common sense" ideas to the discussion which all need to be challenged in order to proceed.

Interdisciplinarity is also seen by several scientists as a process that involves significant input from experts in different domains of knowledge. It could also refer to a network of projects in different domains towards a common goal. Through communication, mutual skill sharing, and collaboration, experts from various domains would find a creative solution to a difficult problem that could not be answered by any individual subset of fields on its own. Other scientists also had a more nuanced understanding of multidisciplinarity vw. Interdisciplinarity ("fruit salad" vs. "smoothie"), with interdisciplinarity requiring a deeper interaction and synthesis of research efforts. Here, problems, perspectives, concepts, theories, data, techniques and everything else are integrated in order to advance the scholars' understanding of the studied phenomenon. These efforts often result in a new terminology and paradigm -sometimes even a new discipline itself.

*2. How would you measure interdisciplinarity in your project?*
This is a question that not many scientists have thought about or they think this is a difficult, or maybe even impossible metric. When measurable, interdisciplinarity can be evaluated both qualitatively and/or quantitatively. The responses to this survey show that some meaningful ways in which interdisciplinarity can be measured for the Astro2020 community are:
- Diversity of competencies necessary to answer a specific question or address a specific problem;
- Years of training, time in a field;
- The extent of a project in questions, methods, literature, etc.
- Departmental affiliation or nature of degrees of major authors of papers.
- Bibliometric tools (journal categories classified by Web of Science), both for papers and for researchers (*"A researcher or a team publishing in journals that have been in different categories or in a journal that has more than one category indicates that their research is interdisciplinary."*)
- Keyword network map (network science algorithms)
- Shannon's entropy, an index to measure the diversity of a habitat in ecology, to measure the disciplinary diversity (i.e., NASA Astrobiology Institute's teams and projects.)

*3. What is your main field/s and/or topic/s of research, either current or past?*
Most scientists that have replied to this question have "moved" through different disciplines and projects, with backgrounds in science and technology but pursuing sociological topics or with backgrounds in social sciences but pursuing science specific topics, with a wide breadth of research topics.

*4. If your field IS NOT in the social sciences and/or humanities, what do you expect the most from a collaboration with the social sciences (including history, sociology, anthropology, economics, etc.,…; and including topics such as gender, race, diversity issues, workforce development, collaboration, etc….). How would this be most impactful for your research?*
The non-social scientists pointed to the need of diversity of fields for better solving problems, for challenging their current assumptions and for identifying best practices for recruiting/retaining a diverse workforce, from undergraduates on up, as well as identifying barriers to success.

*5. If your field IS in the social sciences and/or humanities, how do you envision a collaboration with NASA? What would your input and research be most useful for?*
The social scientists pointed to the need of STEM education for underrepresented groups and young girls, improving critical thinking skills, the accurate understanding of misuse of social sciences in projects (i.e., do not "use" social science, but integrate social scientists in teams), including social scientists into crafting research agendas, asking comprehensive research questions, facilitating collaboration and communication among team members, including social scientists into proposal development, researching societal impacts of science and technology, communicating science with the public.

*6. If you had a grant from NASA for your current research, how would you use social science? Who would you invite to be a part of your team and with what goals?*
This question was answered with the following ideas: the variety of uses to which the term "life" is applied, create a bio-ontology team to specifically address useful ways to construct, formalize, and communicate categories related to life, position papers on 1) the level of current consensus among astrobiologists, 2) the concrete effect categories have on research and mission priorities, 3) the concrete effect of categories on communication with the public, 4) recommendations for achieving consensus, and 5) recommendations for useful categories, reconsidering the "why" of space exploration in order to clarify the goals and build an interdisciplinary ethical framework for human engagements with space, how to best support students for success.

*7. If you could ask NASA Headquarters about social sciences and their impact within the organization, either administratively or research oriented, what would your question/s be?*
The questions the scientists would like to ask are numerous, and some of them are mentioned here:
What additional resources and people would be needed to do said research justice? With these collaborations, how does one bring in people from additional interdisciplinary fields? And how does one incorporate such researchers – as consultants, staff researchers or other- and what does the mechanism then look like? What fields are currently represented in the community? In NASA funded activities? How much integration occurs? At what stages of planning? At what stages of career development? What goals does current leadership have? What is known about the effectiveness of past events at achieving those goals? Why is "human factors" a distinct unit focusing on the technical problems of human space exploration rather than an aspect of all research programs which requires incorporating social scientific expertise in every project? Have you considered that every project/lab should have a social scientist assigned to it? Questions about the metrics for interdisciplinarity, novelty, productivity, etc., that are either measured in a primitive way or not measured at all and not included in the evaluation process. How to measure? How to integrate? How to evaluate?

8. *What kind of research activities do you think would be useful in this space?*
The wide range of answers on this mentions gender and women in STEM issues that address the problems early on, disabilities inclusions, motivations for life research, studies on effective communication, community formation, studies of language usage in astrobiology, studies about diversity of academic disciplines, nationalities, economic background, studies on effective governance and best strategies for group decision making, studies on effective funding mechanisms, How have conceptual structures shaped mission architecture (including but not limited to implicit definitions of life)? How has the text authorization bills (for NASA, ESA…) and treaties (OSU…) been shaped by conceptual and ethical structures? (e.g., JAXA's mission is particularly informative.) How have those texts influenced the work done? How have expectations for non-Earth life changed through time? How have those expectations shaped scientific research programs and specific experiments?, studies on better interaction with experts from risk assessment and popular culture that could help us better understand the impacts of how we communicate our work searching for life elsewhere in the universe[1], Who is talking the most during meetings, who is interrupted and who does the interrupting, who has more access to resources, who has a nicer office, who meets with the boss the most.

9. *Can you give some examples of papers that you used in your research that did interdisciplinary work between social sciences and other sciences?*
There were many responses with literature in this space and this is now captured in the references list of this white paper, as well as in the key points of this paper.

10. *Are you aware of any other organizations or professional associations that did interdisciplinary studies (i.e., NIH, NSF, NIST, etc….)? If so, can you mention them briefly?*
The following institutions have conducted interdisciplinary studies: ISHPSSB (https://www.ishpssb.org/), SSoCIA (http://ssocia.com/), AAAS DoSER (https://www.aaas.org/programs/dialogue-science-ethics-and-religion/about), IAU and IAC on interdisciplinarity of SETI, Horizon2020 and Horizon Europe, NIH on the science of teams, Science of Team Science Network, and the Association for Interdisciplinary Studies

11. *In your professional opinion, why do we or do we not use social sciences as often as we should? ("should" being a very subjective statement here)?*
The community is recognizing the faulty division between "soft" and "hard" sciences or "natural" and "social" sciences, that is now starting to get blurred. Some mention this as normal scientific siloing and stereotypical intellectual arrogance. One of the reasons behind this is the cultural difference between these natural and social sciences, which includes different language (jargon), goals, standards of evidence, and expectations for consensus. Additionally, both social sciences and humanities have rejected strong statements of disciplinary agreement, making it difficult to identify recognized experts, and both Social Sciences and Humanities represent and encompass huge umbrellas of disciplines, topics and fields. Astrobiology is set apart as it has attracted a low, but significant, level of interest from historians of science, philosophers of science, ethicists (especially environmental), theologians, sociologists, and anthropologists. It is unclear how rising experts in those fields fit into the astrobiology community. It is also unclear how they contribute (wish to contribute, are asked to contribute) to common identity and goals. This makes it hard to see, judge, or encourage their commitments and it can also discourage long-term participation. Another reason is that scientists think they are objective as an inherent subconscious bias and forget they are human (i.e. subjective by nature).

---

[1] Discussions of Mars as a habitat for humanity and "Earth 2.0" or "Earth twins" lead people to believe it is less important that we preserve the Earth's biosphere, which is an unintentional and potentially catastrophic consequence of that communication.

*12. What would you like social scientist to measure in your preferred topics and fields of research?*
The answers here have been quite specific: How many people are working in "astrobiology"? How do they identify in terms of academic discipline? What are the primary journals? How many articles do they publish? What reach do those articles have? What other journals do the same authors publish in? Who attends astrobiology conferences? What other conferences have astrobiology sessions? Who, outside NASA, are recognized members of mission teams? Which assumptions in our field are likely on shaky ground? Can they help us communicate our research to the public more responsibly?

*13. Are you aware of any data sources regarding field interdisciplinarity and/or interdisciplinary success rates and if yes, can you link that here?*
The surprising answer here has been more or less "no", with the exception of some sparse resources such as Science of Team Science Network, Association for Interdisciplinary Studies.

**IV. Strategic Plan**

**6. Barriers to Effective Interdisciplinary Work including Social Scientists**
One of the largest barriers to interdisciplinary work is time. It takes time to effectively learn what a scientific community does and make a contribution. Another barrier is education and training, together with career options. Currently there are unclear training pipelines or career options for social scientists who want to do this type of interdisciplinary work. And of course, the third but not the least largest barrier is funding for the social scientists, if we are working in support of the science e.g. on the implications of the work the scientists do, *funding should come from the science agencies.* Social science funding (i.e. the competitions social scientists usually use) is often more devoted to other numerous and serious social issues such as domestic violence, refugee integration, income inequality, etc. Normal conference fees for the social sciences are much lower, while conferences such as IAC or space sciences conferences are out of reach for most social scientists. (Denning & Dick, 2019).

**7. Paths Forward**
Based on this community work and effort and the direct interactions with scientists on both social sciences and other sciences, some of the immediate paths forward can be regular seminar series, workshops, ongoing conversations from which collaborations can grow, funding support for travel and collaborations, visiting scholarships and invitations, inclusion of social scientists into the projects. More details about what the community wants and needs can be found above in the answers to this informal survey. And, last for this section, funding and conducting a research project with a formal, wide community survey and interactions with all scientists in these fields, not only in Astro2020 decadal, can shed light into the most pressing and most important interdisciplinarity with social sciences that is needed today.